\documentclass[final,12pt]{elsarticle}

\usepackage{amssymb}
\usepackage[T1]{fontenc} 
\usepackage[utf8]{inputenc}

\journal{ArXiv}

\begin{document}

\begin{frontmatter}



\title{Low frequency Raman study of the Boson peak in a Tellurite-tungstate glass over temperature}


\author[mymainaddress,mysecondaryaddress,my3address]{Marcos Paulo Belançon\corref{mycorrespondingauthor}}
\cortext[mycorrespondingauthor]{Corresponding author}
\ead{marcosbelancon@utfpr.edu.br}

\author[mysecondaryaddress]{Guilhem Simon}

\address[mymainaddress]{Universidade Tecnológica Federal do Paraná\\Câmpus Pato Branco -  Grupo de Física de Materiais\\CEP 85503-390, Via do conhecimento Km 01, Pato Branco, Paraná, Brazil}
\address[mysecondaryaddress]{Université Pierre-Marie Curie - Paris VI\\Câmpus Jussieu - Laboratoire Monaris (UMR 8233)\\Place Jussieu Tour 43-53 Case Courrier 49 75252 PARIS CEDEX 05, France}
\address[my3address]{Universidade Estadual de Campinas\\Optics and Photonics Research Center (CEPOF)\\Rua Sérgio Buarque de Holanda 777, CEP 13083-859, Campinas, SP, Brazil}

\begin{abstract}

In this work we present our findings performing low frequency Raman measurements over temperature in a tellurite-tungstate glass sample. The spectra could be fitted by using only two log-Gaussians, which suggests that quasi-elastic light scattering is negligible. The positions of the peaks have the same temperature behavior, being the higher frequency one refered as Boson Peak in literature. The similar temperature behavior and depolarization ratio indicates that the peaks may have the same origin, being linked to the tranversal and longitudinal vibrations of some unit.

\end{abstract}


\end{frontmatter}


\section{Introduction}
\label{intro}

Glasses have been widely exploited by mankind since milennia ago and nowadays those materials are keys for many technologies. The photonics world market for example, which is growind faster than the world economy, employes glasses in lasers, fibers, amplifiers, and many other devices. This have put the interaction between light and matter in evidence in the last few years, aiming to enhance properties, increase efficiency and develop new technologies for many strategic and fundamental fields, such as energy\cite{Huang2013} and communication\cite{Desurvire2011,Ballato2013}.

Even though glasses have been so extensively used and studied, some properties remains a puzzle. One of the most interesting example is the observed population of states at Terahertz frequencies found in glasses and any amorphous material\cite{Hong2008,Hou2015,Mallamace2015}. This excess of vibrational density of states at such frequencies is known as ``Boson Peak'' (BP) and it has been observed since the 70's\cite{Shuker1970} in Raman spectra, however, its origin is still subject of debate in the literature\cite{Franz2015,Hyeon-Deuk2016,Shen2016}. 

To ilustrate the discussion we can look for example for the model of Martin and Brenig\cite{brenig1974}, by which the BP should consist of both transversal and longitudinal mode acoustic waves, due to a short scale correlation range over mechanical and electrical properties in non-crystalline materials. This could permit phonons of wavelength matching this scale to travel without the usual attenuation of the desordered network. Another important contribution was given by Malinovsky and Sokolov\cite{Malinovsky1986}, by showing that the form of the BP is independent of the chemical composition and thermal prehistory. By this way, some universal property of desordered materials is expected to describe the BP origin. However, on the recent literature\cite{Shintani2008} we may found evidences that only tranversal phonons (not the longitudinal ones) should be universally linked to the BP.

Looking for evidences of such short scale units in literature, one may found for example the systematic studie about the BP performed by Nemanich\cite{Nemanich1977}. He found in some chalcogenide glasses structural units with a size compatible to the Martin and Brenig model, however, this were not confirmed in all glasses in the study. Buchenau\cite{Buchenau1986} have studied the origin of BP in silica glass by neutron difraction, and the findings have suggested that such populations of states at BP frequencies are originated by the rotation of $SiO_4$ tetrahedra. Duval\cite{Duval1986} had confirmed the existence of ``aggregates'' or ``microcrystallites'' matching the size related to the BP observed in Raman spectra of a silicate glass. Among other evidences of the existence of such units inside the glass, we may point the observation\cite{Umverslty1988} of a new BP by nucleating microcrystallites into the material. Some authors have called this units by ``blobs''\cite{Duval1999,Schroeder2004,Hong2008}.  However, a few works have shown that the nature of BP may be more complicated\cite{Masciovecchio1996,Monaco2009,aiChumakov2011} that only the ``blobs'' based model.

In the low energy range of Raman spectra Quasi-elastic light scattering (QES) may also be expected\cite{To1968,Fallis2013a,Malinovsky2000,Fontana2005,Zanatta2011a,Takahashi2012,Osipov2015} at this low energy range. In liquids the QES originates from the Doppler shift of the light reflected by one moving particle with speed $v$ relative to the detector, and this phenomena is the basis of a few experimental techniques, such as laser doppler velocimetry. In solids however, the QES origin is still on debate\cite{Fontana2005} and it seems to be a concensus that this band is centered at $0cm^{-1}$ and may extent until $\sim50cm^{-1}$\cite{Takahashi2012} in glasses. 

Experimental data available in the literature shows us that BP position usually ranges from $\sim40cm^{-1}$ to $\sim80cm^{-1}$\cite{Schroeder2004}, depending on composition. The BP Temperature dependence have been investigated in Silica\cite{Fontana2005}, Calcium-aluminosilicate\cite{Schroeder2004}, Germania\cite{Zanatta2011a}, Phosphate\cite{Takahashi2012}, Borate\cite{Osipov2015}, Tellurite-Zinc\cite{Stavrou2010} and Tellurite-Oxyhalide\cite{Kalampounias2007} glasses, and the results agree that increasing temperature red shifts the peak. This last observation does also agree with the ``blobs'' idea, once that increasing temperature the material is expanding, which is in favor of longer wavelength phonons, i.e. a shift towards lower frequency. On the other hand, pressure induces a blue shift\cite{Schroeder2004,Hong2008,Weigel2016}. All these experimental findings have drived some work looking for a correlation between fragility and BP\cite{Astrath2006,Mallamace2015}.

One can conclude that the informations available on literature are still not the enough to clarify the question. If we are thinking in more experimental work, one should note that acquire Raman shift data at this range and below can be very difficult. To mention a few points, it is well known that Raman shift intensity may be more or less intense in each sample, in such way that the noise/signal ratio may change our ability to measure ultra low frequencies in some glasses. Once we are usually working in the limit of sensitivity and resolution, just a few $cm^{-1}$ of the laser line, and taking into account that the relative intensity between BP and QES (or even a third band) are different from one sample to another, seldom deconvolute and compare these spectra may be challenging.

By looking to the discussion in the literature, the models still being developed and tested\cite{Milkus2016a,Marruzzo2013,Chamberlin2013} or considering that the better understanding of BP could result in the exploration of new phenomena\cite{Luo2016} and even in improvements of glass transparency\cite{Champagnon2009}, we have performed the low frequency Raman study presented here. The sample investigated is a tellurite-tungstate glass (in \%molar $71TeO_2-22.5WO_3-5Na_2O-1.5Nb_2O_5$, which we will name here as TWNN) produced by the melting quenching process under controlled atmosphere as described elsewhere\cite{Belancon2014a}. Two peaks instead of the only one usually found in literature were observed at frequencies $<45cm^{-1}$. By deconvoluting them, their temperature behavior in the range $30^oC$ to $460^oC$ could be obtained and them discussed.

\section{Methods}

\begin{figure}[h]
\center
\includegraphics[scale=0.4]{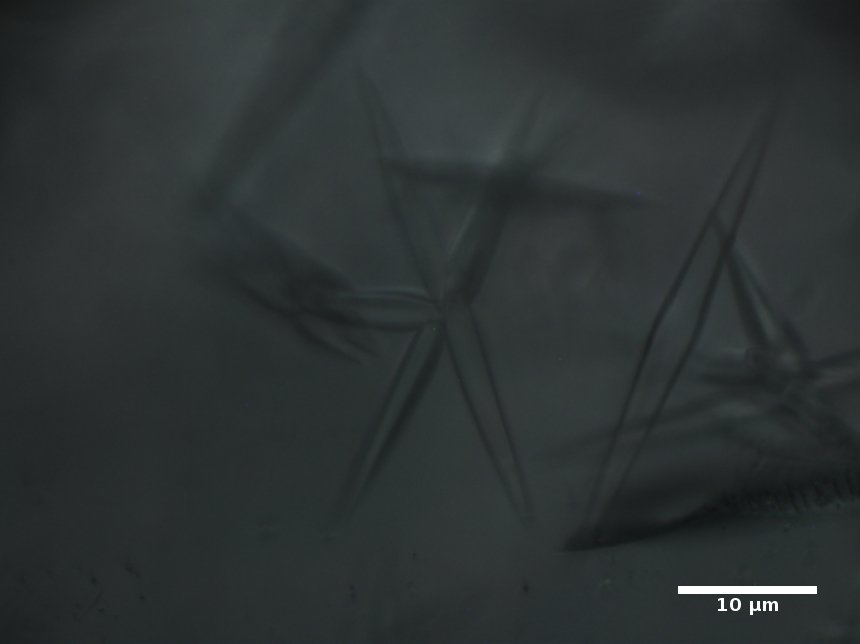}
\caption{Example of crystal observed growing during the Raman experiments.}
\label{cristal}
\end{figure}

The measurements were performed in a Raman spectrophotometer (Horiba Jobin Yvon HR 800) under excitation power of 37 $mW$ at 514.5 nm (Argon Laser). A x50 microscope objetive lens were used. The sample was heated at steps of $50^oC$, and a shorter step of $10^oC$  between $330^oC$ and $400^oC$ to have a better insight around the glass transition. The measurements were performed 15 minutes after the system had reached each sample, to avoid inaccurate temperature data. If by one hand this helped to be confident about the sample's temperature, by another some crystallization were observed, as we can see in figure \ref{cristal}.

As we have interpreted that this crystallization have induced Raman peaks $\sim60cm^-1$, to deconvolute and compare properly the ultra low frequency peaks observed we have used only the data in the range $3-53cm^{-1}$, which were $\sim100$ data points. All data are presented in log-scale, in which as we are going to show two gaussians could fit pretty well the data. The peak centered at $0cm^{-1}$, usually is atributted to QES and fitted by a Lorentzian\cite{Osipov2015} or power law\cite{Stavrou2010} function. Here we have just keeped the data for $\omega<3cm^{-1}$ out of the fitting process, and the results justify by itself the empyrical analysis performed. 

\section{Results and Discussion}

\begin{figure}[h]
\includegraphics[scale=0.5]{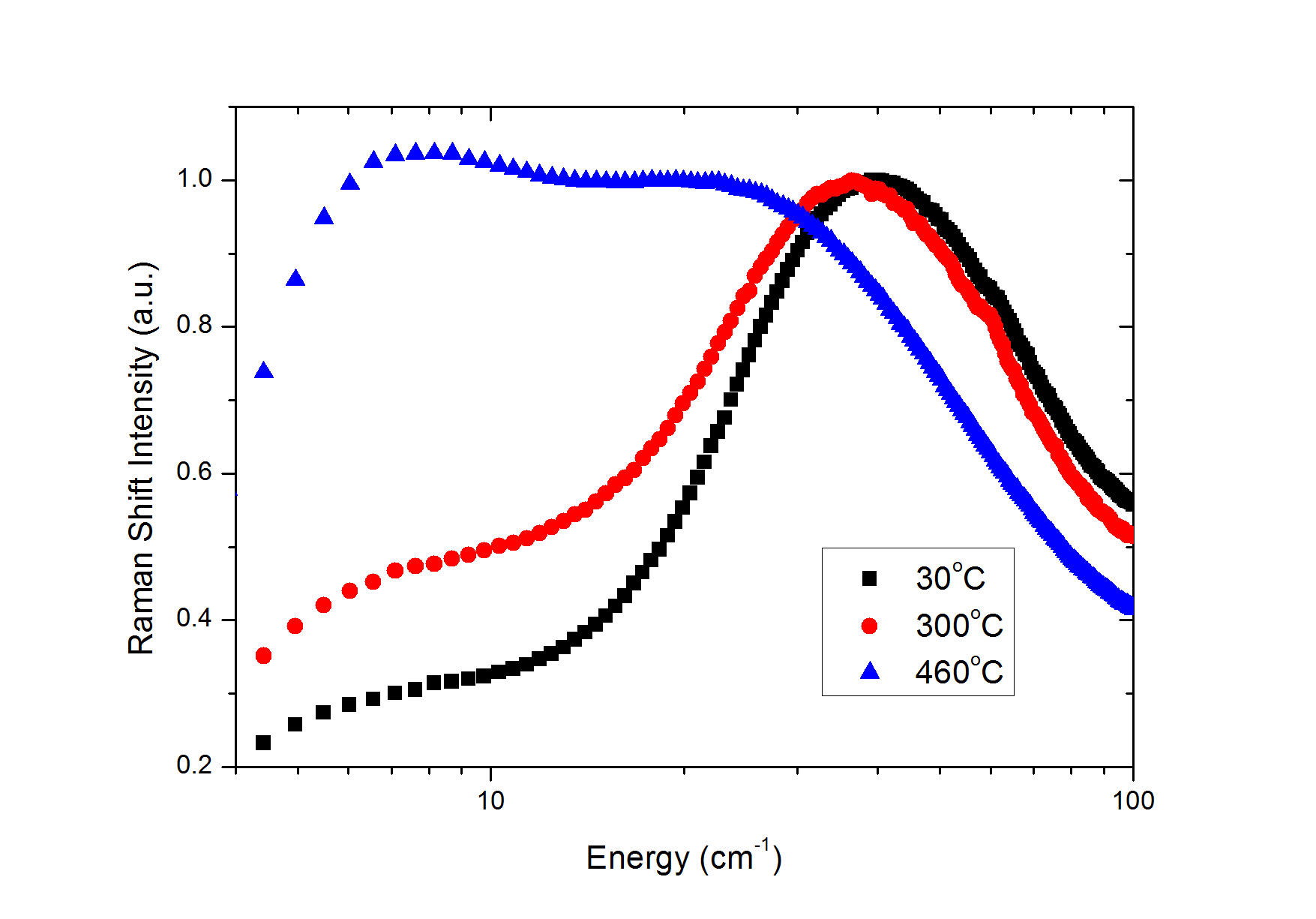}
\caption{Low frequency Raman spectra. Once that at $~60cm^{-1}$ the crystallization produces a new peak during heating, in all spectra just the range $3-53cm^{-1}$ was selected to be fitted.}
\label{example}
\end{figure}

Figure \ref{example} shows the low frequency Raman spectra for room temperature, before glass transition $(T=300^oC)$ and for the highest temperature reached in our experiments$(T=460^oC)$. As one can see, the characteristic shape of BP can be clearly seen at $\sim40cm^{-1}$, which is the same position where it is found in other Tellurite glasses\cite{Murugan2005,Kalampounias2007,Gebavi2009,Stavrou2010}. However, the shape of the curves around $3-12cm^-1$ are for the best of our knowledge particularly different from the data for other materials.

\begin{figure}[h]
\includegraphics[scale=0.5]{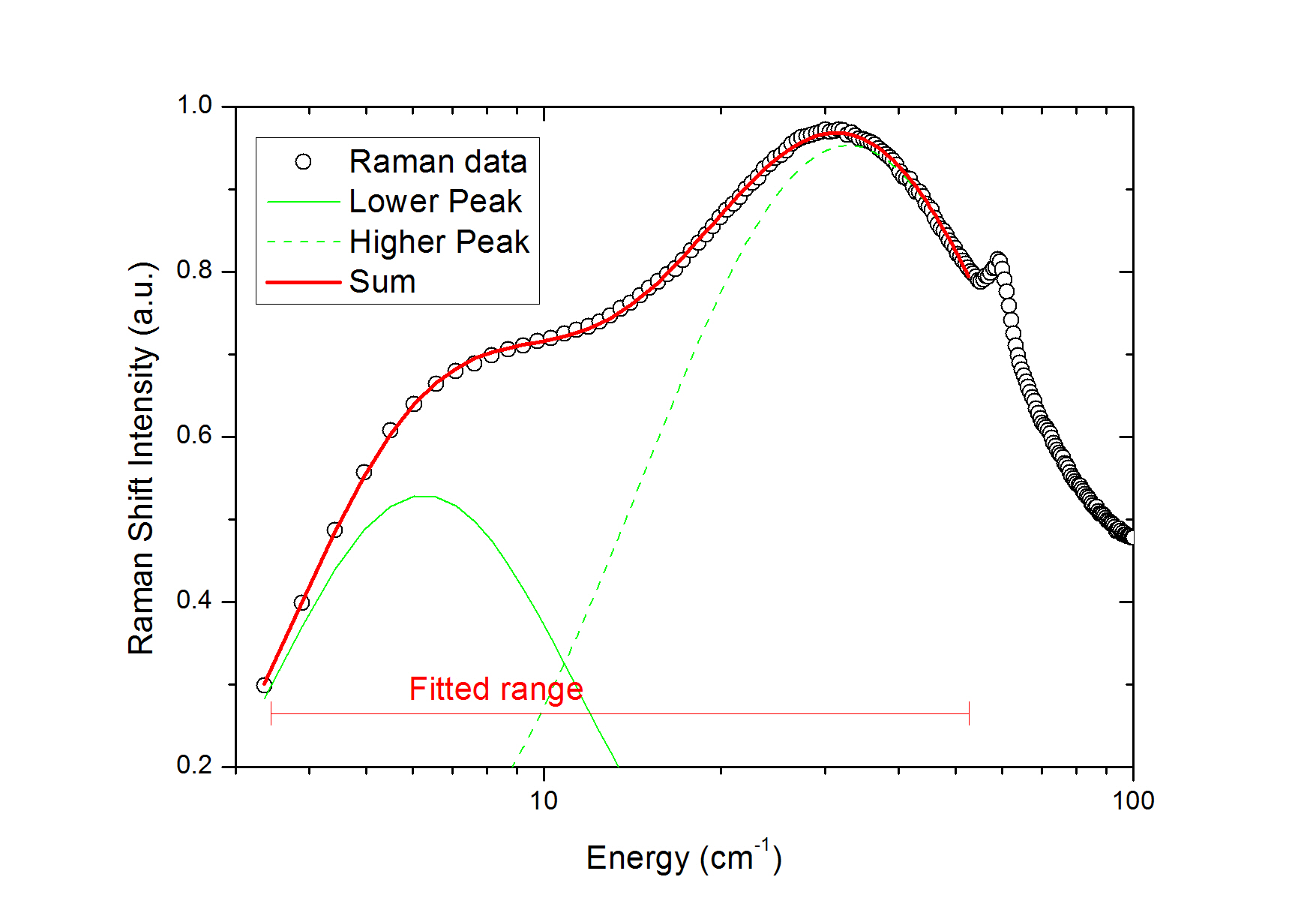}
\caption{Raman spectrum at $420^oC$ plotted to show an example of the fitting process.  Once that at $~60cm^{-1}$ the crystallization produces a new peak during heating, in all spectra just the range $3-53cm^{-1}$ was selected to be fitted.}
\label{first}
\end{figure}

\begin{figure}[h]
\includegraphics[scale=0.5]{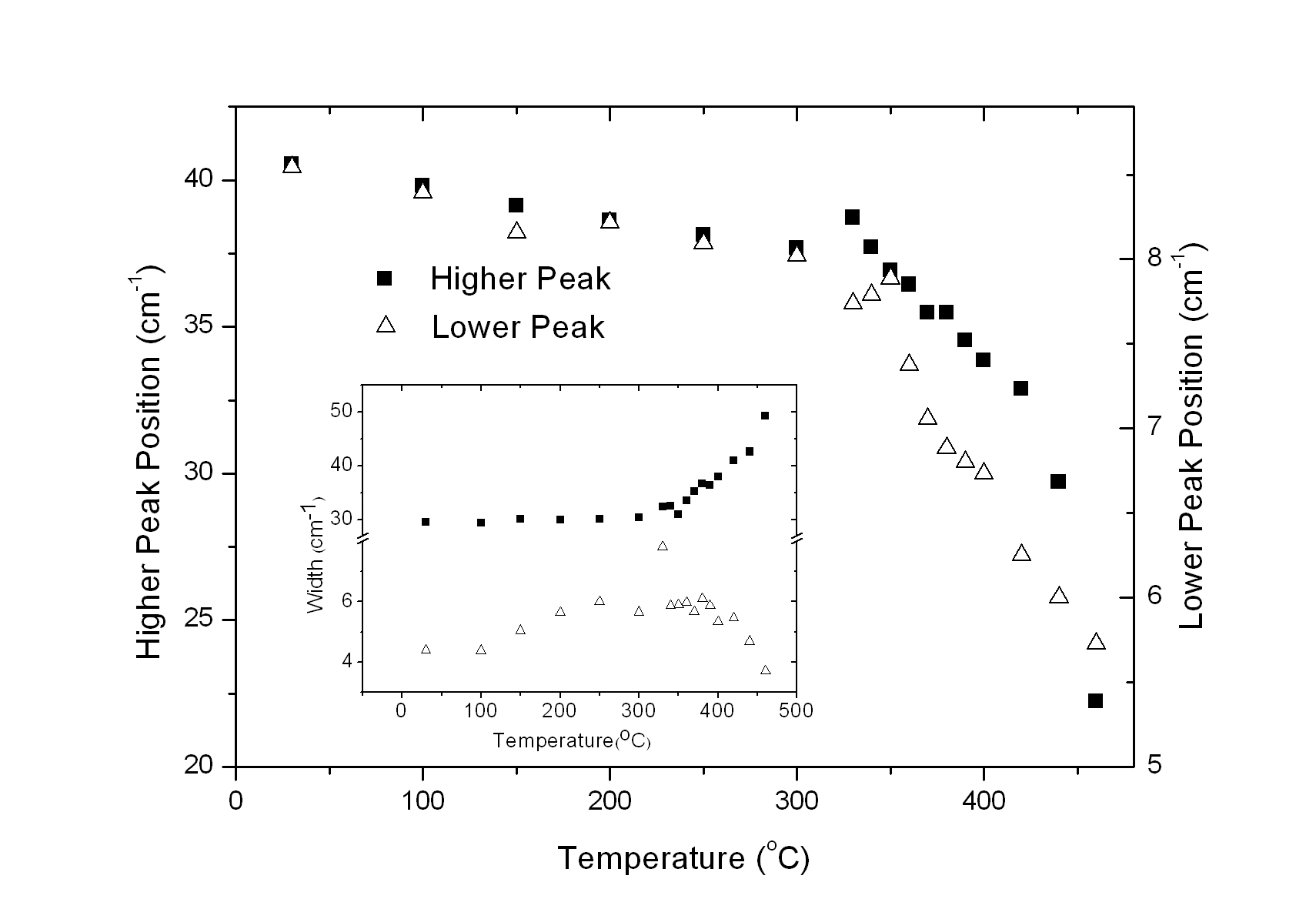}
\caption{Temperature dependence of BP frequency for both peaks (higher peak scaled at left and lower peak at right). As one can see they have very similar behaviors, more clearly observed before the glass transition at ~$350^oC$.}
\label{second}
\end{figure}

To explain our analysis, in figure \ref{first} we show the fitting result for the spectrum at $420^oC$. As one can see, the data can be deconvoluted in two peaks centered at $\sim6cm^-1$ and $\sim30cm^{-1}$, respectively. We found remarkably that these two peaks could fit the data in such way, giving for all spectra a coefficient of determination $R^2>0.999$. This may indicate that the intensity ratio QES/BP is neglible in our sample at this range. The figure \ref{second} shows the position of both peaks as function of temperature, as well the width of them which are plotted in the inset.

Both peak positions are decreasing linearly until $T_g$ (which is $\sim300^oC$), matching the same straight line if we took a normalized scale. The point for us is not if the peaks have one or another dependence with temperature, but that both have the same dependence and by this way we are motivated to conclude that they are produced by the same mechanism. On the other hand, the data on the inset suggests a nearly temperature independent width for both peaks before the glass transition, while the higher peak becomes broader and the lower peak, may be, have a tendence to become thinner after $T_g$. It should be pointed that we do not have too much data points for the lower peak, and it is located just a few $cm^-1$ of the laser line, where QES usually is important. By this way, our fitting process is less accurate for the lower peak and also by the fact that QES and BP intensity may not have the same temperature dependence. This means that QES may start to be more significant at some temperature, and the lower peak should be more affected than the higher peak.

Taking the data until $300^oC$, and normalizing each one by their maxima to have the relative change of BP position over temperature, we found the values of $(-2.2\pm0.4)10^{-4}/^oC$ and $(-2.6\pm0.1)10^{-4}/^oC$ for the angular coefficient for the lower and higher peaks, respectively. Our interpretation for this similarity in the temperature dependence is that the peaks are correlated to same mechanism. By the model of Martin and Brenig, as mentioned in the introduction the BP should consist of both transversal and longitudinal acoustic waves. On this point of view, if a ``blob'' is about one wavelength of such waves we should obtain something like the two peaks observed here; each one being related to the longitudinal and tranversal acoustic waves, respectively.

\begin{figure}[h]
\includegraphics[scale=0.5]{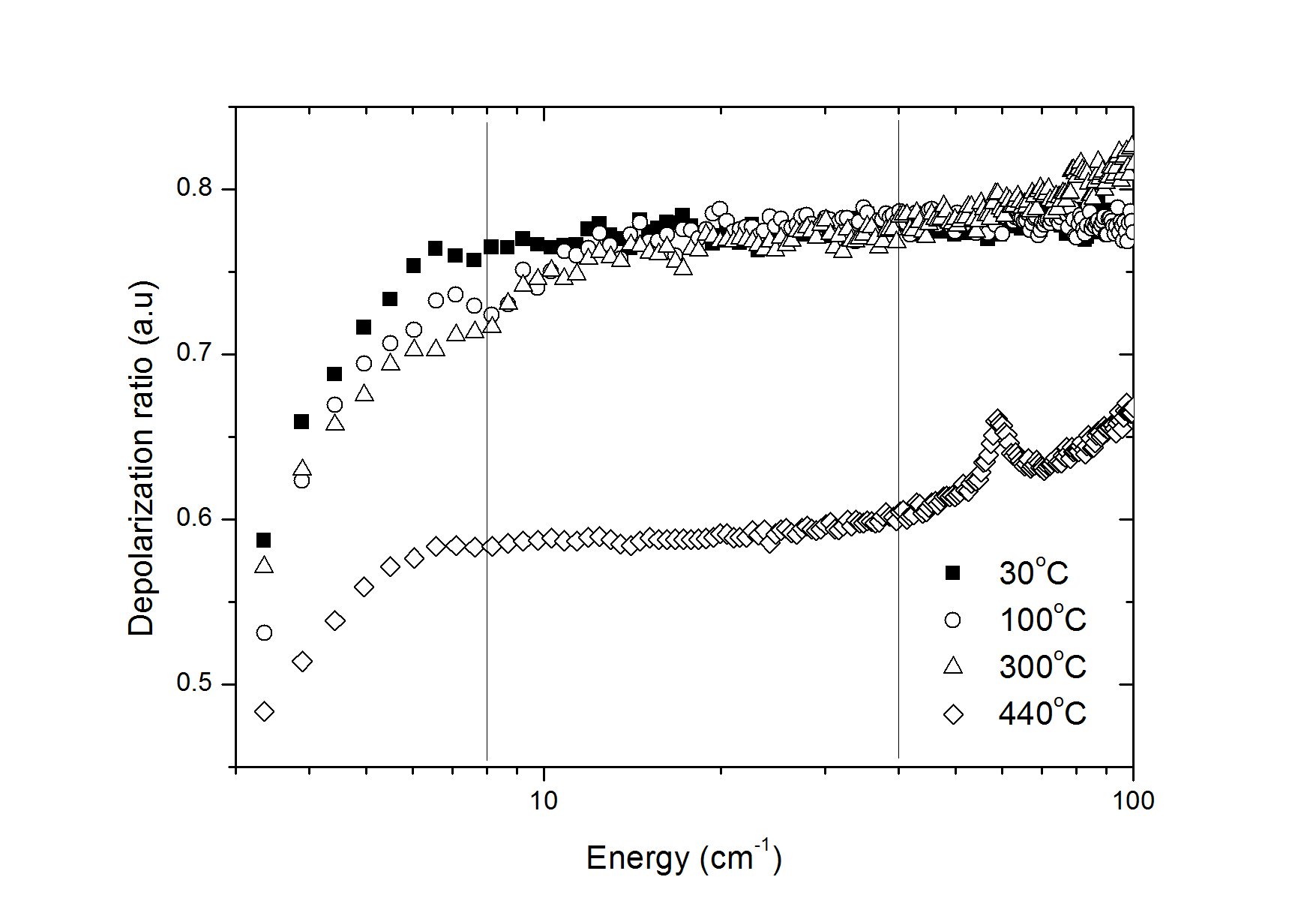}
\caption{Depolarization ratio for selected temperatures.}
\label{five}
\end{figure}

In figure \ref{five} we show the depolarization ratio in the same range of the Raman spectra shown before. The two vertical lines indicates the regions where the two peaks were found in this work. As one can see, the depolarization remains around $0.75$ at room temperature in the range $6-60cm^{-1}$. Heating the sample this ratio has decreased, initially in the lower peak range where at $100^oC$ we have got $0.72-0.73$ against $0.75-0.77$ for the higher peak range. However, for the temperatures above the glass transition, as shown here for the $440_oC$, the depolarization is nearly frequency independent, exhibiting a shape very similar to that observed at room temperature.

\section{Conclusion}

Our findings can be summarized as follows: 1) TWNN glass exhibits an unusual low frequency Raman spectra, 2) which can be deconvoluted by a pretty simple two gaussian fit. 3) The peaks observed have the same temperature dependence and 4) depolarization ratio, even though the depolarization dependence over temperature could be slightly different. 5) QES, if thougth as a band centered at $0cm^{-1}$ seems to be negligible or at least very weak if compared to the BP bands in our TWNN sample. By this way, the results shown here can be interpreted in terms of the Martin and Brenig model, where each peak is the result of tranversal and longitudinal vibrations of a ``blob''.

In order to study the puzzle of BP in glasses, we have performed this low frequency Raman investigation. Our data analysis is pretty simple and fully experimental, i.e. we fitted the data and described their behavior. Our findings may contribute in the development and discussion about the BP models, as well motivate more work by Raman and other techniques to enrich the experimental information available and drive the community towards clarify the mechanisms behind the energy states in the low frequency range in amorphous materials.

\section*{Acknowledgement}

The authors would like to thank CNPq (INCT/FOTONICOM, grants 574017/2008; 480576/2013-0 and 504677/2013-6) and FAPESP (grants 2005/ 51689-2, 2008/57857-2 and 2012/04291-7) for their financial support. A special thanks to Dr. Christophe Petit, the director of the Monaris laboratory during the execution of this study.




  \bibliographystyle{elsarticle-num}
  \bibliography{/home/mbelancon/Documentos/library}






\end{document}